\documentstyle[sprocl]{article}

\input{psfig.sty}

\def\be{\begin{equation}}
\def\ee{\end{equation}}
\def\bea{\begin{eqnarray}}
\def\eea{\end{eqnarray}}

\begin{document}

\title{THE $\gamma \to 3\pi$ FORM FACTOR AS A CONSTRAINT ON\\ 
SCHWINGER--DYSON MODELING OF LIGHT QUARKS}

\author{DUBRAVKO KLABU\v CAR}

\address{Physics Department, P.M.F.,
   Zagreb University, Bijeni\v{c}ka c. 32, Zagreb, Croatia}

\author{BOJAN BISTROVI\' C}

\address{ Center for Theoretical Physics, Laboratory for
Nuclear Science and Department of \\ Physics, 
Massachusetts Institute of Technology, Cambridge, Massachusetts 02139  }

\maketitle\abstracts{ 
The form factor for $\gamma \pi^+ \to \pi^+ \pi^0$ was calculated 
in a simple--minded constituent model with a 
constant quark mass parameter, as well as in the Schwinger-Dyson 
approach.
The comparative discussion of these and various other
theoretical results on this anomalous process, 
as well as the scarce already available data
(hopefully to be supplemented by more accurate CEBAF data soon),
seem to favor Schwinger--Dyson modeling which would 
yield relatively small low--momentum values of
the constituent (dynamically dressed) quark mass function.
}

\begin{figure}[t]
\psfig{figure=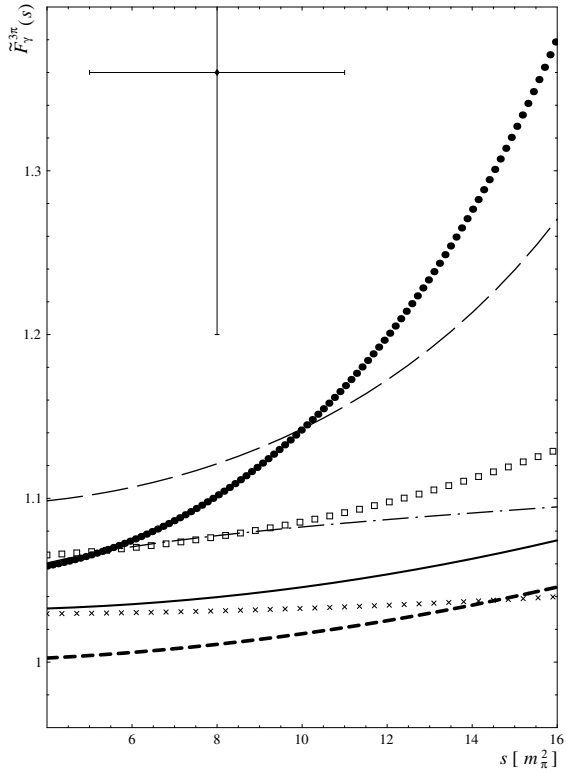,height=4.1in}
\caption{Various predictions for the dependence of the normalized
$\gamma 3\pi$ form factor ${\widetilde{F}}^{3\pi}_\gamma$ on the 
Mandelstam variable $s\equiv (p_1 + p_2)^2$. The kinematics is as 
in the Serpukhov measurement: the photon and all three pions are 
on shell, $q^2=0$ and $p_1^2=p_2^2=p_3^2=m_\pi^2$.  \label{fig:figure}}
\end{figure}

\noindent The form factor for the anomalous process 
$\gamma(q)\to \pi^+(p_1) \pi^0(p_2) \pi^-(p_3)$ 
was calculated as the quark ``box"-amplitude 
in two related approaches \cite{BiKl99PRD,BiKl9912452}. 
In our Ref. \cite{BiKl99PRD}, the intermediate fermion ``box" loop is 
the one of constituent quarks with the constant quark mass parameter $M$.
The predictions of this quark loop model \cite{BiKl99PRD} are given
in Fig. 1 by the long-dashed curve for $M=330 \, \rm{MeV}$,
by the line of empty boxes for $M=400 \, \rm{MeV}$, and by
the line of crosses for the large value $M=580 \, \rm{MeV}$.
(In the lowest order in pion interactions, they are also the 
form factors of the $\sigma$-model and of the chiral quark model.)
Our second Ref. \cite{BiKl9912452} employs the Schwinger-Dyson (SD) 
approach \cite{Roberts0007054}, 
where the box loop amplitude is evaluated with the
dressed quark propagator 
\begin{equation}
S(k)= \frac{1}{i \rlap{$k$}/ \,A(k^2) + m + B(k^2)}
       \equiv \frac{Z(k^2)}{i \, \rlap{$k$}/ \, + {\cal M}(k^2)}
\label{EuclS}
\end{equation}
containing the {\em momentum-dependent},
mostly dynamically generated quark mass function ${\cal M}(k^2)$, 
while $m$ is the {\em small} explicit chiral symmetry breaking.
In Fig. 1, the solid curve gives our $\gamma 3\pi$ form factor obtained 
in the SD approach for the empirical pion mass, $m_\pi = 138.5$ MeV,
while the dashed curve gives it in the chiral limit, $m_\pi = 0 = m$. 
To understand the relationship between the predictions
of these two approaches, one should, besides the curves in Fig. 1,
compare also the analytic expressions we derived for the 
form factors [esp. Eqs. (20)--(21) in Ref. \cite{BiKl9912452}
and analogous formulas in Ref. \cite{BiKl99PRD}]. 
This way, one can see, first, why the constant, 
momentum-independent term is smaller in the SD case, causing 
the downward shift of the SD form factors with respect to those 
in the constant constituent mass case. Second, this constant term 
in the both approaches diminishes with the increase of the 
pertinent mass scales, namely $M$ in the constant-mass case, 
and the scale which rules the SD--modeling and which is of course 
closely related to the resulting scale of the {\em dynamically generated} 
constituent mass ${\cal M}(k^2\sim 0)$. Finally, the 
momentum--dependent terms are similar in the both approaches;
notably, the coefficients of the momentum expansions (in powers
of $p_i \cdot p_j$) are similarly suppressed by powers of
their pertinent scales. This all implies a
transparent relationship between ${\cal M}(k^2)$ at small $k^2$
and the $\gamma 3\pi$ form factor, so that the accurate CEBAF data,
which hopefully are to appear soon \cite{Miskimen+al94}, 
should be able to constrain ${\cal M}(k^2)$ at small $k^2$,
and thus the whole infrared SD modeling. 
Admittedly, we used the Ball--Chiu Ansatz for the dressed quark--photon 
vertex, 
but this is adequate since Ref. \cite{MarisTandy9910033} found that for 
$-0.4 \, {\rm GeV}^2 < q^2 < 0.2 \, {\rm GeV}^2$, the true solution 
for the dressed vertex is approximated well by this Ansatz, plus 
the vector--meson resonant contributions which however vanish 
in our case of the real photon, $q^2=0$.
Therefore, 
if the experimental form factor is measured with sufficient precision 
to judge the present SD model results definitely too low, it will
be a clear signal that the SD modeling should be reformulated
and refitted so that it is governed by a smaller mass scale and 
smaller values of ${\cal M}(k^2\sim 0)$.

The only already available data,
the Serpukhov experimental point \cite{Antipov+al87} (shown in the upper
left corner of Fig. 1), is higher than all theoretical predictions
and is probably an overestimate. However, the SD predictions are
farthest from it. 
Indeed, in the momentum interval shown in Fig 1, the SD form factors
are lower than those of other theoretical approaches (for reasonable
values of their parameters) including vector meson dominance \cite{Rudaz84}
(the dotted curve) and of chiral perturbation theory \cite{Holstein96}
(the dash-dotted curve).
Therefore, even the present experimental and theoretical 
knowledge indicates that the momentum--dependent mass function 
in the SD model we adopted \cite{BiKl9912452}, 
may already be too large at small $k^2$, where its typical value for 
light $u, d$ quarks is ${\cal M}(k^2\approx 0) \approx 360$ MeV. 
Note that some other (so far very successful, 
see review \cite{Roberts0007054}) SD models obtain even 
significantly higher values, ${\cal M}(k^2\approx 0)\approx 600$ MeV
and more,
which would lead to even lower $\gamma 3\pi$ transition form factors.

\section*{Acknowledgments}
D. Klabu\v car thanks the organizers, W. Lucha and N. Brambilla, for 
their hospitality at the Fourth International Conference on ``Quark 
Confinement and the Hadron Spectrum" in Vienna, 3.-8. July 2000.

\end{document}